 \documentclass[final,5p,times,twocolumn]{elsarticle}

\usepackage{amssymb}
\usepackage{lipsum}
\usepackage{amsmath}
\usepackage{natbib}

 \usepackage{lineno}

\makeatletter
\renewcommand\section{\@startsection{section}{1}{\z@}%
  {-18\p@ \@plus -4\p@ \@minus -4\p@}%
  {12\p@ \@plus 4\p@ \@minus 4\p@}%
  {\normalfont\large\bfseries\raggedright}}
\renewcommand\subsection{\@startsection{subsection}{2}{\z@}%
  {-18\p@ \@plus -4\p@ \@minus -4\p@}%
  {8\p@ \@plus 4\p@ \@minus 4\p@}%
  {\normalfont\normalsize\bfseries\raggedright}}
\makeatother

\biboptions{sort&compress}
 
\journal{Physics Letters B}

\begin{document}

\begin{frontmatter}




\title{Baryon-to-Meson Ratios in Jets from Au+Au and $p$+$p$ Collisions at $\sqrt{s_{\mathrm{NN}}} = 200$ GeV }


\author{The STAR Collaboration}

         

\begin{abstract}
\label{abstract}
Jet probes have been used extensively to gain insights into QGP properties, with substantial modifications to jet yields and internal structures seen across multiple measurements, compared against $p$+$p$ results. 
Despite apparent medium-induced changes to jet fragmentation patterns and strong modifications of particle production at intermediate momenta ($2.0 < p_{\rm{T}} < 5.0$ GeV/$c$) in heavy ion collisions, RHIC hadron-hadron correlation results indicate that jet-related baryon-to-meson ratios remain similar to those of $p$+$p$ measurements and are significantly different from those of the QGP bulk. To look for possible medium effects on jet fragments at RHIC, we employ jet-track correlation and particle identification to perform the first measurement of in-jet proton-to-pion yield ratios for charged-hadrons with $2.0 < p_{\rm{T}}  < 5.0$ GeV/$c$. We present the first in-cone (within a radial distance of $R$ from the determined jet axis direction) baryon-to-meson yield ratios associated with reconstructed charged-particle jets from Au+Au and $p$+$p$ collisions at $\sqrt{s_{\mathrm{NN}}} = 200$ GeV using the STAR detector at RHIC. The measured in-jet ratios are found to be consistent within uncertainties between the two systems for the selected kinematic regime, despite significant differences between inclusive ratios for the same systems.
\end{abstract}




\end{frontmatter}



\section{Introduction}
\label{introduction}
Heavy-ion collisions provide a unique environment for studying the Quark-Gluon Plasma (QGP), an exotic phase of matter which consists of deconfined quarks and gluons, in a laboratory setting. QGP properties can be studied by comparing heavy-ion collisions to $p$+$p$ collisions, in which QGP is not expected to be formed. Some key signatures of QGP observed through such comparisons include significant modification of inclusive charged-particle spectra, jet quenching, and enhancement of relative baryon-to-meson production~\cite{BRAHMS_WhitePaper_2005,PHOBOS_WhitePaper_2005,STAR_WhitePaper_2005,PHENIX_WhitePaper_2005,RAA}. Jets, collimated collections of particles produced by fragmentation and hadronization of hard-scattered partons, are present in both $p$+$p$ and heavy-ion collisions. This makes them ideal in-situ probes for studying the QGP, as we can observe the differences in jet properties with and without the presence of the medium. Jet-related phenomena can be studied with fully reconstructed jets or, by proxy, through hadron production measurements at high transverse momenta, $p_{\rm{T}}$. Employing these observables, many measurements at the LHC and RHIC have demonstrated that jets are indeed modified in the presence of the QGP~\cite{JetQuench, Awayside}. 

In contrast with elementary collision systems, hard scattering is not the dominant source of particle production in heavy-ion collisions at intermediate transverse momenta $(2.0 < p_{\rm{T}} < 5.0$ GeV/$c)$, as demonstrated by an enhancement in baryon production relative to meson production~\cite{spectra}.
This enhancement is attributed to strong hydrodynamic flow and coalescence of partons from the medium \cite{Fries_recombination_2003, Greco_coalescence_2003}. It remains unclear to what extent hard-scattered partons contribute to in-medium coalescence processes, and if the QGP presence modifies the resulting in-jet hadrochemistry at RHIC energies. Earlier RHIC measurements have probed jet-associated baryon-to-meson ratios via dihadron correlations, employing high-$p_{\rm{T}}$ hadron triggers as a proxy for jets. A measurement by PHENIX of hadron-triggered jet-like correlations showed that the baryon-to-meson ratio of away side jet-associated hadrons increases with centrality and $p_{\rm{T}}$, suggesting that jet-medium interaction may modify jet-associated hadrochemistry \cite{PHENIX_awayside}. In contrast, STAR measurements of near-side jet-like correlations with identified strange hadrons found no baryon enhancement in the near-side jet-like component, with the measured $\Lambda/K^0_S$ ratio remaining comparable to that in $p$+$p$ \cite{STAR_strange}. Recent simulations from A Multi-Phase Transport model (AMPT) predict modifications to the baryon-to-meson ratio in jets~\cite{ampt}. This study presents the first search for modification in in-jet hadrochemistry at RHIC by combining jet-track correlation with the excellent particle identification (PID) available at STAR to measure the proton-to-pion ($p/\pi$) ratio in jets for both $p$+$p$ and $0 - 10 \%$ central Au+Au collisions at center-of-mass energy per nucleon pair, $\sqrt{s_{\mathrm{NN}}} = 200$ GeV. Deviation between in-jet $p/\pi$ ratios in Au+Au and $p$+$p$ collisions, if it were observed, may be indicative of QGP medium effects.

The distance between a charged-particle track and the jet axis is defined as $\Delta r = \sqrt{(\Delta\phi)^{2}+(\Delta\eta)^{2}}$. The AMPT simulations predict a qualitative upward linear trend in the $p/\pi$ ratio as a function of $\Delta r$ in jets from heavy-ion collisions as compared to the same ratio in $p$+$p$ collisions \cite{ampt}. Although the AMPT predictions were calculated for Pb+Pb collisions at $\sqrt{s_{\rm{NN}}} = 5.02$ TeV, the mechanisms driving the predicted trends, parton energy loss and coalescence hadronization, are expected to be present in Au+Au collisions at $\sqrt{s_{\rm{NN}}} = 200$ GeV as well. Thus, a measurement of the in-jet $p/\pi$ ratio as a function of $\Delta r$ in heavy-ion collisions at RHIC energies would serve to further search for evidence of shower-thermal recombination.
\section{Analysis Procedure}
\label{Methods}
\renewcommand \thesubsection{\roman{subsection}}
\begin{figure*}
	\centering 
	\includegraphics[width=0.9\textwidth, angle=0]{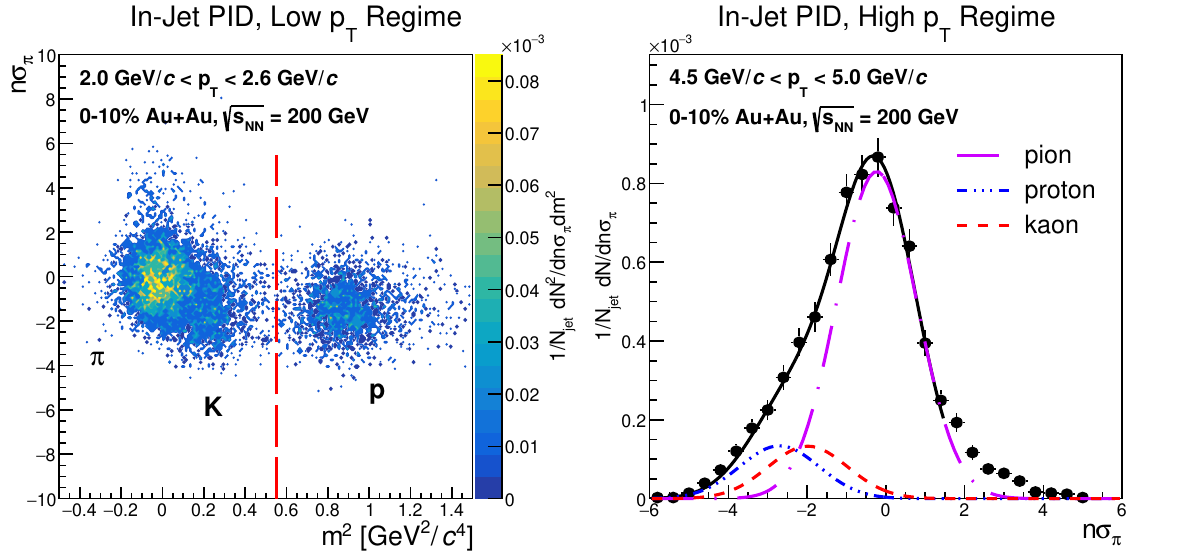}	
	\caption{(left) In-jet 2D PID distribution with $2.0$ GeV/$c < p_{\rm{T}} < 2.6$ GeV/$c$. the per-jet charged-particle yield is shown as a function of both $m^{2}$ ($x$-axis) and $n\sigma_{\pi}$($y$-axis). A red line is overlayed at the $m^{2}$-minimum used as a cut for proton identification in the low $p_{\rm{T}}$ regime. (right) in-jet $n\sigma_{\pi}$ distribution for $4.5$ GeV/$c < p_{\rm{T}} < 5.0$ GeV/$c$. A triple gaussian fit of the three particle species is superimposed over the per-jet charged-particle yield. Both plots represent jets from $0-10\%$ central Au+Au collisions at $\sqrt{s_{\rm{NN}}}=200$ GeV.} 
	\label{PID}%
\end{figure*}
%
The analysis used $\sqrt{s_{\rm{NN}}} = 200$~GeV Au+Au and $p$+$p$ data samples collected by STAR in 2014 and 2015, respectively. A Barrel Electromagnetic Calorimeter (BEMC)~\cite{BEMCNIM} high tower trigger requiring a transverse energy of at least $3.4$ GeV deposited in at least one tower is employed to provide a sample enriched with jets. We analyze 20 million ($0-10\%$ central) Au+Au events and 12 million p+p events. A sample of $1$ million minimum-bias Au+Au events is also employed in our combinatorial background correction. Charged-particle tracks are reconstructed in the Time Projection Chamber (TPC)~\cite{TPC_NIM}. Charged-particle tracks employed in this analysis are required to have a minimum of 25 TPC hits, $N_{\rm{hits}}$, for quality of track reconstruction. The primary vertex is determined algorithmically by fitting the trajectories of charged-particle tracks from the TPC back to a common point~\cite{vzz}. The primary vertex, $V_{z}$ is required to be within $|V_z|<25$ cm. The difference between reconstructed vertex and vertex determined via Vertex Position Detector (VPD)~\cite{VPD}, $V_{VPD}$, is required to be within $|V_z - V_{VPD}|<3$ cm, as the VPD provides an independent validation of the vertex determined by the TPC to minimize out-of-bunch pileup. TPC tracks are required to have a distance of closest approach, $DCA$, to the the reconstructed vertex of $DCA < 1.0$ cm. Event centrality was determined from the charged-particle reference multiplicity measured at mid-rapidity in the STAR TPC and related to collision geometry which is then binned into fractions of  total cross section using a Monte Carlo Glauber approach \cite{MillerGlauber2007}. Particle identification and jet-track correlation is performed identically for both Au+Au and $p$+$p$ data sets.

\indent Particle identification is achieved using information from the TPC and Time of Flight (TOF)~\cite{ToF_NIM} detectors. The TPC measures momentum and energy loss, d$E/$d$x$, and TOF provides the relativistic velocity, $\beta$, of tracks extending from the primary vertex to the TOF detector, which can be used to calculate mass. TOF-matching is required for all TPC tracks considered in the analysis. The two parameters employed are normalized energy loss, $n\sigma_{\pi}$ (Eq.~\ref{nspi}), and mass squared, $m^{2}$ (Eq.~\ref{m2}). $n\sigma_{\pi}$ is defined as 
\begin{equation}
n\sigma_{\pi} = \frac{\ln{(\mathrm{d}E/\mathrm{d}x)_{\rm{measured}}}-\ln{(\mathrm{d}E/\mathrm{d}x)_{\rm{theory}}^{\rm{\pi}}}}{\sigma(\ln{(\mathrm{d}E/\mathrm{d}x)})}, 
\label{nspi}
\end{equation}
where $(\mathrm{d}E/\mathrm{d}x)_{\rm{measured}}$ is the measured energy loss from the TPC,  $(\mathrm{d}E/\mathrm{d}x)_{\rm{theory}}^{\rm{\pi}}$ is the theoretical expectation for the energy loss of a pion, and $\sigma(\ln{(\mathrm{d}E/\mathrm{d}x)})$ is the resolution of the energy loss measurement. Mass squared is measured using track velocity, $\beta$, from TOF as 
\begin{equation}
m^{2} = \frac{p^{2}}{c^{2}}\left(\frac{1}{\beta^2} - 1\right), 
\label{m2}
\end{equation}
where $p$ is the measured charged-particle track momentum. \\
\indent Different PID methods are employed at the low and high $p_{\rm{T}}$ regimes. Two representative $p_{\rm{T}}$ bins are shown in Fig.~\ref{PID}. For $p_{\rm{T}} < 3.0$ GeV/$c$, $m^{2}$ resolution allows the proton yields to be directly summed above a $m^2$-minimum of 0.5 $\rm{GeV}^{2}/\it{c}^{\rm{4}}$, given the clear separation of the proton peak from those of pion and kaon. 
The remaining signal is then projected onto $n\sigma_{\pi}$ and fit with a double Gaussian to extract the pion yield. 
For $p_{\rm{T}} > 3.0$ GeV/$c$, $m^{2}$ resolution deteriorates, meaning the proton yield cannot be directly summed. In this kinematic range, relativistic rise leads to improved separation in d$E/$d$x$ between particle species allowing the full distribution to be fit with a triple Gaussian to extract proton, pion, and kaon yields simultaneously \cite{rise}. The triple Gaussian fit is constrained to $-6.0 < n\sigma_{\pi} <1.5$, to avoid contamination from electrons present in the positive $n\sigma_{\pi}$ regime. A cut on track pseudorapidity, $|\eta_{\rm{track}}| < 1.0$, is applied to all tracks in the analysis, driven by TOF coverage. Particle species-dependent TPC tracking efficiency corrections are applied to the measured proton and pion yields before constructing a $p/\pi$ ratio. TOF-matching efficiency is the same for the two particle species in the selected kinematic regime.

Two distinct track $p_{\rm{T}}$ selections enter this analysis. The first is the constituent-track selection ($p^{\rm{cons}}_{\rm{T}}$), which is used only for jet reconstruction and axis determination. The second is the associated-track selection ($p_{\rm{T}}$), which is used to construct jet-track correlations and to extract identified-particle yields relative to the reconstructed jet axis. The constituent-track selection limits which particles enter the jet finder, while the final in-jet PID measurement is obtained from all charged-particle tracks correlated with the reconstructed jet axis within a selected $\Delta r$ region. This allows the measurement to probe identified-particle production below the constituent threshold while  keeping the background levels manageable for jet reconstruction.

A jet-track correlation technique is employed for the extraction of $p/\pi$ ratios in jets. Jets are reconstructed using the anti-$k_{\rm{T}}$ algorithm~\cite{antikt}, with jet radii $R = 0.2, 0.3, 0.4$ using constituent transverse momentum selection of $p^{\mathrm{cons}}_{\rm{T}} > 2.0$ GeV/$c$, and $R = 0.3$ for $p^{\mathrm{cons}}_{\rm{T}} > 3.0$ GeV/$c$. Jets with uncorrected momenta  $p^{\rm{jet}}_{\rm{T}} > 9.0$ GeV/$c$ 
are considered for the analysis. The jet axis pseudorapidity is limited to $|\eta_{\rm{jet}}| < (1.0-R)$, ensuring that the entire jet cone is contained within the $\eta_{\rm{track}}$ acceptance. For each event, correlations are constructed in $\eta$ and azimuthal angle, $\phi$, for all charged-particle tracks in the event with respect to the jet axis, achieving a distribution in $\Delta\eta$ and $\Delta\phi$, where $\Delta\eta = \eta_{\rm{jet}} - \eta_{\rm{track}}$ and $\Delta\phi = \phi_{\rm{jet}} - \phi_{\rm{track}}$. 

Because both the jet and associated track are restricted to finite $\eta$ ranges, the raw jet-track correlation contains a pair-acceptance structure that must be corrected.
A mixed event (ME) method is employed for this  \cite{CMS_jettrack_2016}. The mixed event distribution is constructed by correlating each track in an event with a jet axis from a different event in our sample. The resulting ME distribution is normalized to unity at maximum, $ME(0,0)$ (corresponding to the maximum pair acceptance), and the signal correlation is divided by the ME. The correction results in an underlying event distribution that is uniform in $\Delta\eta$. The final pair acceptance-corrected charged-particle density-per-trigger jet is found as:

\begin{equation}
\frac{1}{N_{\rm{jet}}}\frac{d^{2}N}{d\Delta \eta d \Delta \phi} = \frac{ME(0,0)}{ME(\Delta \eta, \Delta \phi)} SE(\Delta \eta, \Delta \phi),
\label{acceptance}
\end{equation}
where $N_{\rm{jet}}$ is the number of jets in the sample, $SE(\Delta \eta, \Delta \phi)$ is the same event per-jet normalized jet-track correlation, and $ME(\Delta \eta, \Delta \phi)$ is the mixed event per-jet normalized jet-track correlation. These distributions are given by:

\begin{equation}
\begin{aligned}
SE(\Delta \eta, \Delta \phi) = \frac{1}{N_{\rm{jet}}}\frac{d^{2}N^{\rm{same}}}{d\Delta \eta d \Delta \phi} \\
ME(\Delta \eta, \Delta \phi) = \frac{1}{N_{\rm{jet}}}\frac{d^{2}N^{\rm{mixed}}}{d\Delta \eta d \Delta \phi}.
\end{aligned}
\label{SEME}
\end{equation}


After the pair acceptance correction is applied, a circular region with a radius equal to the jet resolution parameter, $R$, is selected about the jet axis as the jet signal region, and an equal area away from the jet peak in $\Delta\eta$ is selected to estimate the underlying event (UE) contribution. The area selected for the UE consists of two semi-circular regions with radii $R$, shifted from the jet axis direction to $|\Delta\eta| = 1.0$, leaving a relative pseudorapidity gap of at least $0.2$ between the edge of the in-jet signal selection and the UE. The UE selection is constrained to the same $\Delta\phi$ to ensure that variation of hadrochemistry due to the known differences in baryon and meson collective flow patterns~\cite{flow} are accounted for. For jets with $p^{\rm{cons}}_{\rm{T}}>2.0$ GeV/$c$, the UE yield represents $20\%$ of the total in-cone yield for the lowest $p_{\rm{T}}$ bin presented ($2.0 < p_{\rm{T}} < 2.2$ GeV/$c$), regardless of jet $R$ selection, and falls to a negligible contribution by $3.0$ GeV/$c$. For jets with $p^{\rm{cons}}_{\rm{T}}>3.0$ GeV/$c$, the starting point for UE contribution represents $67\%$ of the total in-cone yield, and rapidly falls to $0\%$. Histograms are constructed in $\Delta\phi$, $\Delta\eta$, $m^{2}$, and $n\sigma_{\pi}$ from these selections, and UE is subtracted from the jet signal in all four variables. 

%

UE subtraction eliminates the background contribution to associated charged-particles; however, two forms of influence from the background on jet reconstruction must also be accounted for: lower-momentum jets combined with an upward fluctuation of the underlying event, and fully combinatorial jets from tracks unrelated to a hard scattering. Two complementary methods are employed to measure these contributions: (i) mixed constituent events (MCE) for combinatorial jets, and (ii) pseudo-embedding to account for upward fluctuations of the underlying event. 

An analysis of MCEs is used to identify the background contribution from combinatorial jets, conceptually similar to the mixed event method employed in Ref. \cite{STAR_hjet_2017}. For the sample of mixed constituent events, a distribution of $N_{\rm{tracks}}$-per-event is first constructed based on the Au+Au events selected for the analysis.
For each event multiplicity extracted from this distribution, an MCE is then created by sampling each track selected from a distinct minimum-bias event, thereby minimizing the introduction of tracks from hard processes. The jet reconstruction is performed on the MCEs, accepting jets with two or more constituent tracks. Jet-track correlation and UE subtraction are performed on the resulting distributions, as with the signal. 

Pseudo-embedding is carried out by combining events with jets from $p$+$p$ collisions with an Au+Au MCE. The jetfinder is run on $p$+$p$ events with a lowered threshold of $p^{\rm{jet}}_{\rm{T}} > 2.0$ GeV/$c$ to identify $p$+$p$ jet seeds. Events containing such $p$+$p$ jet seeds are selected to be embedded into an Au+Au MCE. The $p$+$p$ event is randomly downsampled to match the tracking efficiency of Au+Au events before the two events are combined. The jet reconstruction is then performed on a combined $p$+$p$ $\oplus$ MCE event, and if the resulting jet is matched in location to one of the embedded $p$+$p$ jet seeds, the correlation is constructed using all tracks from the Au+Au MCE against the found jet axis. UE subtraction is performed on the resulting distribution identically to how it is performed in real Au+Au events. The resulting correlations represent the upward fluctuations carried into the jet signal from Au+Au UE.

The final hadron-per-jet distributions for combinatorial jets are scaled by the $N_{\rm{jet}}$-per-event rate in MCE divided by the $N_{\rm{jet}}$-per-event rate in the real Au+Au events. For Au+Au jets with constituent $p^{\rm{cons}}_{\rm{T}} > 2.0$ GeV/$c$, and $p^{\rm{jet}}_{\rm{T}} > 9.0$ GeV/$c$, this rate is found to be $9.3\%$ for anti-$k_{\rm{T}}$ $R$ = $0.2$, $27.5\%$ for $R$ = $0.3$, and $52.8\%$ for $R$ = $0.4$. For Au+Au jets with $R$ = $0.3$, constituent $p^{\rm{cons}}_{\rm{T}} > 3.0$ GeV/$c$, and $p^{\rm{jet}}_{\rm{T}} > 9.0$ GeV/$c$, this rate is found to be $3.2\%$. The contributions obtained from both methods are scaled appropriately and subtracted from the signal to obtain the final corrected in-jet distributions. Pseudo-embedding is performed as a correction for exclusively Au+Au results, to correct for the large heavy-ion background which is not present in $p$+$p$ events.



The radial dependence of in-jet particle production is studied for jets with $p^{\rm{cons}}_{\rm{T}} > 2.0$ GeV/$c$, $R = 0.3$, and $p^{\rm{jet}}_{\rm{T}} > 9.0$ GeV/$c$ in both $p$+$p$ and $0-10\%$ central Au+Au collisions. Additionally, this measurement is performed for in-jet tracks with $2.0 < p_{\rm{T}} < 3.0$ GeV/$c$. 
This $p_{\rm{T}}$ selection is made to ensure the cleanest PID for tracks included in the radial analysis, given that below $3$ GeV/$c$, TOF resolution allows for direct bin-counting of proton yields. This is the $p_{\rm{T}}$ regime that would dominate a full $p_{T}$-integrated measurement, given the falling shape of the particle spectra. The selection of $p^{\rm{cons}}_{\rm{T}} > 2.0$ GeV/$c$ is made to bolster statistics of both total jets and particles in the regime of interest. The selection of $R = 0.3$ is made to mitigate the combinatorial background correction while maintaining sufficient in-jet radial coverage.

For the $\Delta r$-dependent analysis, the jet axis definition is changed from the momentum-sum-based (E-scheme) to that of the direction of the highest momentum constituent. This is similar, though not identical, to the so-called Winner-Take-All (WTA) axis~\cite{antikt}. This choice avoids the known deficiency of the E-scheme definition of the anti-$k_{\rm{T}}$ algorithm, which depletes particle yields at the edge of the jet cone at $\Delta r = R$.  For all $\Delta r$-integrated, $p_{\rm{T}}$-dependent $p/\pi$ ratio measurements presented in this work, the choice of the jet axis makes a negligible impact, and the E-scheme axis is employed. 

The $p_{\rm{T}}$-dependent measurements of in-jet $p/\pi$ ratios were performed by integrating charged-hadron yields from background-subtracted jet-track correlations with a range defined by $\Delta r = R$. For $\Delta r$-dependent PID, the same in-jet PID procedure is performed at three integration ranges, $\Delta r = 0.1,0.2,0.3$, keeping all jet selection criteria unchanged. 
Combinatorial and upward fluctuation background evaluation is performed at each integration radius, following the same pseudo-embedding into MCE procedure described earlier in this section, and subtracted from the in-jet signal before reporting final in-jet identified-particle yield ratios.


Systematic uncertainties due to $n\sigma_{\pi}$ calibration, $m^{2}$ minimum selection for proton identification, and underlying event subtraction are evaluated by running the entire analysis with variations in cuts and fit parameters. Differences are taken between variations and nominal values to evaluate absolute systematic uncertainty independently for each source. The dominant systematic source on the in-jet $p/\pi$ ratios comes from variations in PID. Uncertainty from PID contributes an absolute varation of $\sim0.05$ for jets with $R=0.2$, $\sim0.06$ for jets with $R=0.3$, and $\sim0.1$ for jets with $R=0.4$. The increase in background subtraction for jets with larger radius results in an amplification of systematic uncertainty on the final background-subtracted in-jet $p/\pi$ ratios. Variation in combinatorial background subtraction resulting from pseudo-embedding is also evaluated for all Au+Au in-jet ratios, contributing an absolute systematic uncertainty of $0.025$ on the lowest $p_{\rm{T}}$ bin reported, and smoothly reducing to zero by the highest $p_{\rm{T}}$ bin reported. Systematic uncertainty due to tracking efficiency cancels out in the reported ratio.


\begin{figure*}
	\centering 
	\includegraphics[width=\textwidth, angle=0]{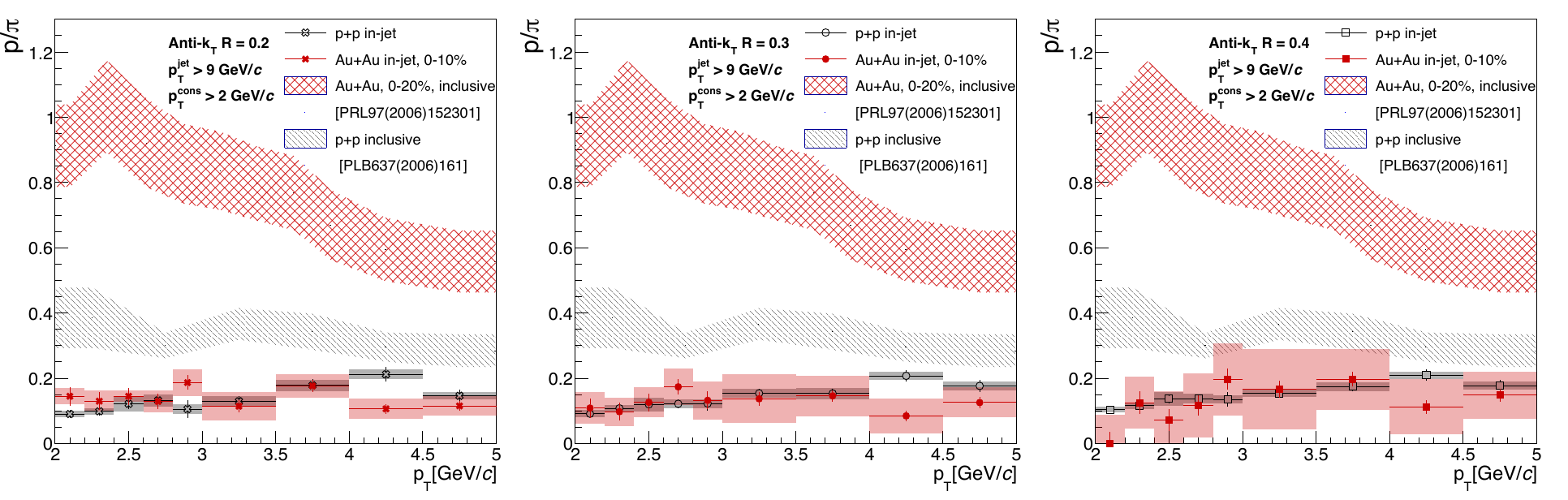}	
	\caption{In-jet $p/\pi$ ratios for $p$+$p$ and $0-10\%$ central Au+Au collisions, with $p^{\rm{jet}}_{\rm{T}} > 9$ GeV/$c$ and $p^{\rm{cons}}_{\rm{T}}$ $> 2.0$ GeV/$c$. Published inclusive $p/\pi$ ratios are included for both systems~\cite{spectra, pp}. Inclusive hadron $p/\pi$ ratios are presented with weak-decay feed-down corrections to the measured pion spectra. In-jet measurements are presented without correction for weak-decay feed-down. Three different jet $R$ parameters are presented: (left) $R = 0.2$, (center) $R = 0.3$, (right) $R = 0.4$.} 
	\label{R_depends}%
\end{figure*}

\begin{figure}[ht]
	\centering 
	\includegraphics[width=0.4\textwidth, angle=0]{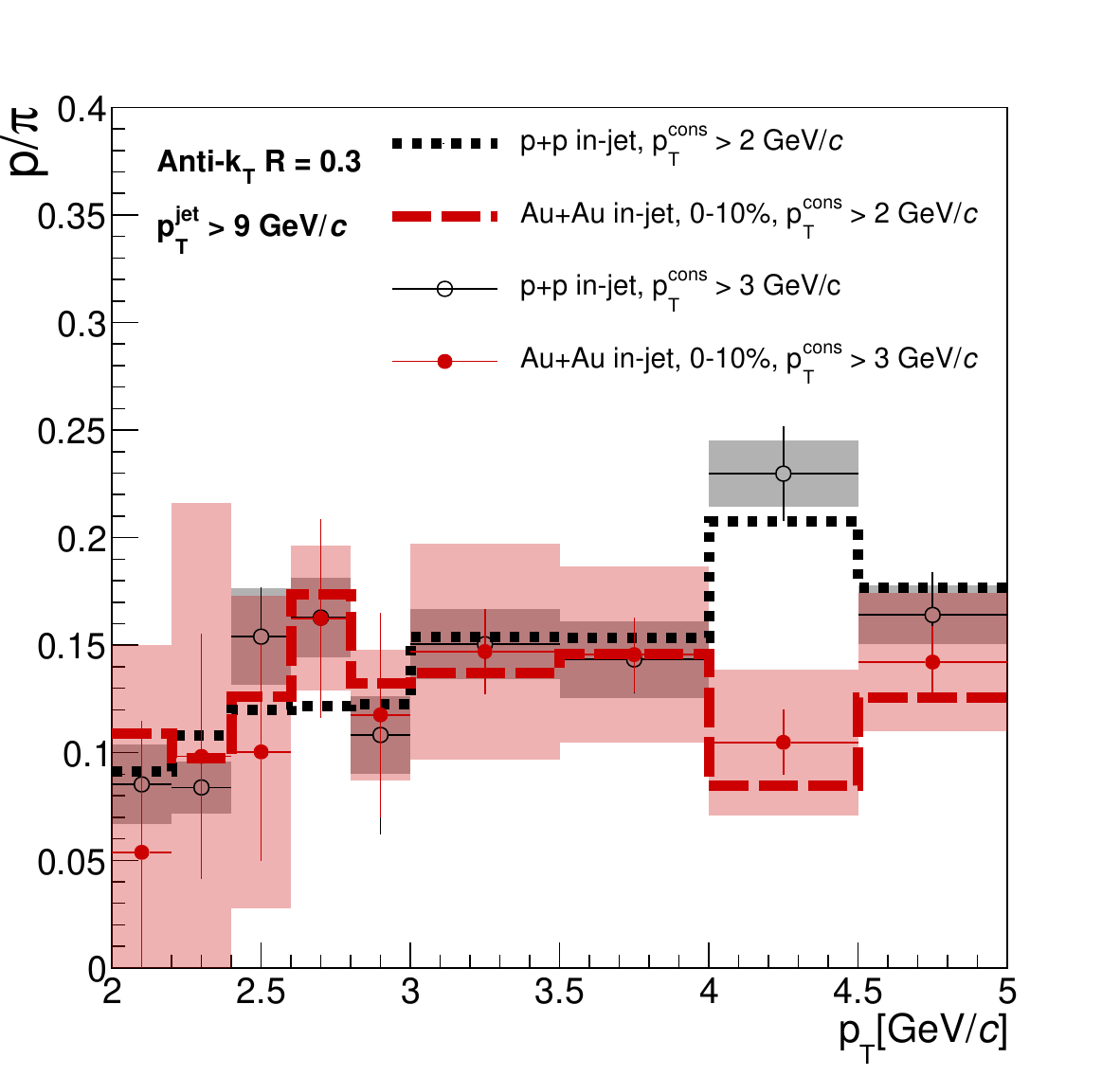}	
	\caption{In-jet $p/\pi$ ratios for $p$+$p$ and $0-10\%$ central Au+Au collisions, with $p^{\rm{jet}}_{\rm{T}} > 9$ GeV/$c$ and $p^{\rm{cons}}_{\rm{T}}$ $> 3.0$ GeV/$c$. Comparison against dashed lines representing in-jet ratios for jets with $p^{\rm{cons}}_{\rm{T}}$ $> 2.0$ GeV/$c$, as presented in Fig.~\ref{R_depends} (center).} 
	\label{Result}%
\end{figure}

\begin{figure}[ht]
	\centering 
	\includegraphics[width=0.4\textwidth, angle=0]{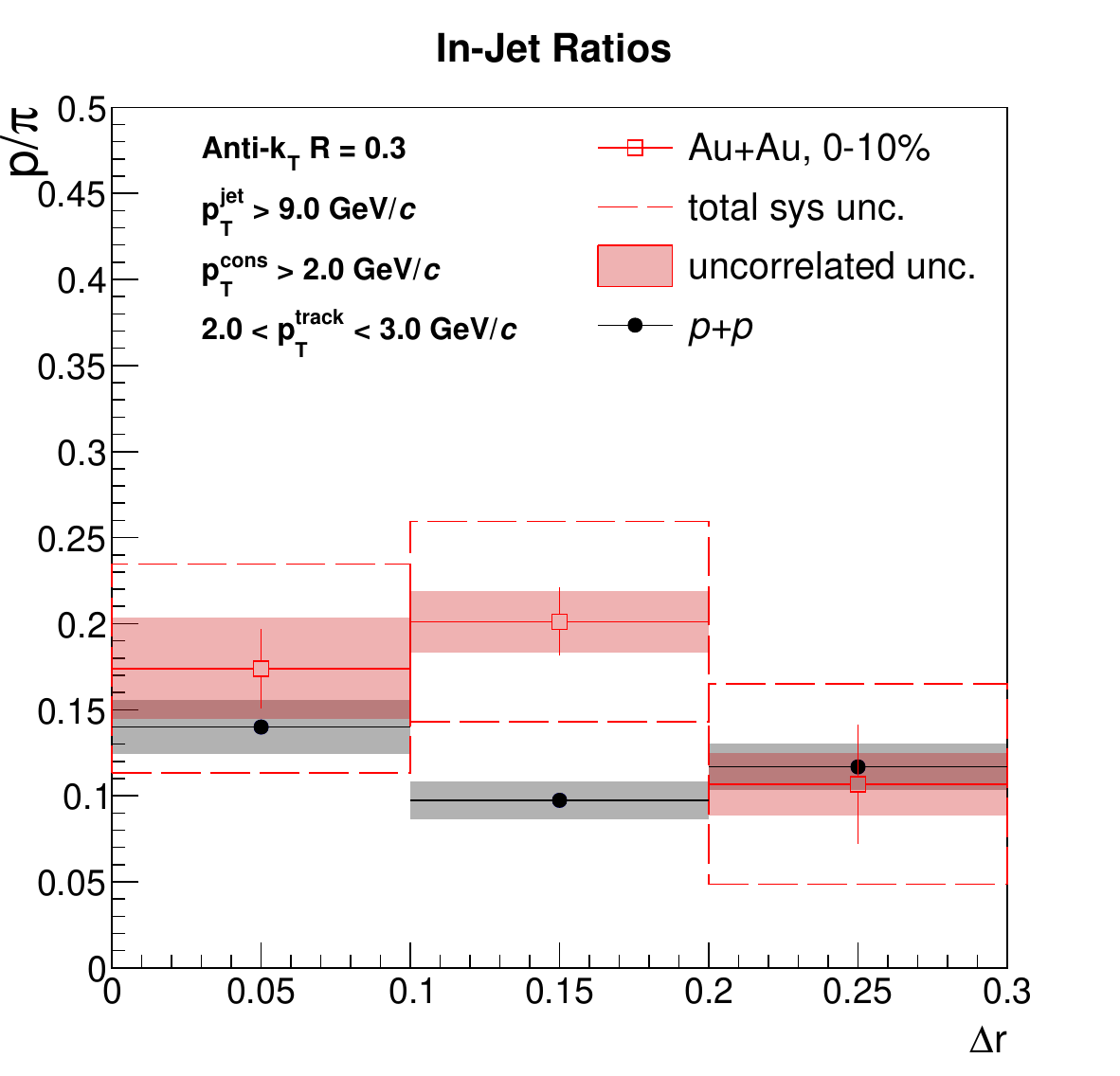}
	\caption{In-jet $p/\pi$ ratios as a function of $\Delta r$ for $p$+$p$ and $0-10\%$ central Au+Au collisions, with $p^{\rm{jet}}_{\rm{T}} > 9$ GeV/$c$ and $p^{\rm{cons}}_{\rm{T}}$ $> 2.0$ GeV/$c$.} 
	\label{dR_fig}%
\end{figure}

\section{Results and Discussion}

\label{RandD}

\begin{table}[htbp]
\centering
\begin{tabular}{ccccc}
\hline
System & $R$ & $\langle p/\pi \rangle$ & Stat. & Sys. \\
\hline
$p$+$p$ & 0.2 & 0.125 & 0.006 & 0.004 \\
Au+Au   & 0.2 & 0.133 & 0.007 & 0.010 \\
\hline
$p$+$p$ & 0.3 & 0.127 & 0.004 & 0.004 \\
Au+Au   & 0.3 & 0.124 & 0.007 & 0.017 \\
\hline
$p$+$p$ & 0.4 & 0.138 & 0.004 & 0.004 \\
Au+Au   & 0.4 & 0.137 & 0.009 & 0.030 \\
\hline
\end{tabular}
\caption{Average $p/\pi$ ratio over $2.0$ GeV/$c <p_{\rm{T}}<5.0 $ GeV$/c$ and uncertainties for each collision system and jet radius $R$.}
\label{average_values}
\end{table}

The in-jet $p/\pi$ ratio from $p$+$p$ collisions at $\sqrt{s} = 200$ GeV is shown in Fig.~\ref{R_depends} for jets with $p^{\rm{cons}}_{\rm{T}}$ $> 2.0$ GeV/$c$, $p^{\rm{jet}}_{\rm{T}} > 9$ GeV/$c$, and three anti-$k_{\rm{T}}$ jet radii $R = 0.2, 0.3, 0.4$. Average $p/\pi$ ratios for all radii are listed in Table \ref{average_values}. For all three jet radii, the observed in-jet ratio is below that measured for inclusive hadrons in $p$+$p$ collisions~\cite{pp}. The in-jet results are presented without correction for weak-decay feed-down. We note that the inclusive pion measurements from $p$+$p$ collisions were corrected for weak-decay feed-down. The correction is on the order of $4\%$~\cite{pp}, and does not qualitatively impact this comparison.
This feature is not unexpected, as data from ALICE \cite{ALICE_ppPID_7TeV} at the LHC show that the inclusive hadron particle production cannot be described in a purely vacuum-like fragmentation framework, instead requiring Color Reconnection (CR) at the partonic level to describe the full particle production~\cite{CR}. In this analysis, the in-jet selection biases the yield towards fragmentation from hard scattering, removing the low $p_{\rm{T}}$ baryonic enhancement contributions from CR. 


The in-jet $p/\pi$ ratios from 0-10\% central Au+Au collisions at $\sqrt{s_{\mathrm{NN}}} = 200$ GeV with $p^{\rm{cons}}_{\rm{T}} > 2.0$ GeV/$c$, $p^{\rm{jet}}_{\rm{T}} > 9.0$ GeV/$c$, and different $R$ are also shown in Fig.~\ref{R_depends}, together with the previously published inclusive hadron $p/\pi$ ratio for $0-20\%$ central events from the same system and collision energy~\cite{spectra}. The in-jet results are presented without correction for weak-decay feed-down. We note that the inclusive pion measurements from Au+Au collisions were corrected for weak-decay feed-down. The correction is on the order of $5\%$~\cite{spectra}, and does not qualitatively impact this comparison. Larger radii are studied to select a sample of jets that could carry a greater level of medium interaction. 
The measured Au+Au in-jet $p/\pi$ ratios show no significant deviations from the $p$+$p$ baseline for the same jet selection criteria, indicating no significant medium effects on the in-jet hadrochemistry for the jets studied. Average in-jet $p/\pi$ ratios over the studied $p_{\rm{T}}$ range for all three jet $R$ selections in both systems are listed in Table.\ref{average_values}, alongside statistical and systematic uncertainties on these averages. No dependence on jet radius is observed. The observed agreement between in-jet $p/\pi$ ratios from the two systems contrasts with the significant enhancement seen in the Au+Au inclusive hadron $p/\pi$ ratio measurement as compared to the inclusive hadron $p/\pi$ ratio from $p$+$p$ collisions. 


In-jet $p/\pi$ ratios are also presented for jets with $p^{\rm{cons}}_{\rm{T}}$ $> 3.0$ GeV/$c$, $p^{\rm{jet}}_{\rm{T}} > 9.0$ GeV/$c$, and anti-$k_{\rm{T}}$ jet radius $R = 0.3$. A higher $p^{\rm{cons}}_{\rm{T}}$ minimum ensures a cleaner sample of jets, reducing the contribution from combinatorial background from $27.5\%$ to $3.2\%$, which significantly reduces the corresponding contribution to the systematic uncertainty. 
This selection of jets with a narrow anti-$k_{\rm{T}}$ radius and a high $p^{\rm{cons}}_{\rm{T}}$ minimum is biased toward harder fragmenting in the event collection. The collection of jets in central heavy-ion data may be subject to ‘survivor bias’, whereby the selected jet population is preferentially composed of jets that have undergone minimal interaction with the medium. This bias can reduce sensitivity to medium-induced modifications and should be considered when interpreting the results. A comparison of this result against the $p^{\rm{cons}}_{\rm{T}}$ $> 2.0$ GeV/$c$ result with $R = 0.3$ is shown in Fig. \ref{Result}. The in-jet $p/\pi$ ratios for the two $p^{\rm{cons}}_{\rm{T}}$ thresholds are in agreement for both $p$+$p$ and $0-10\%$ central Au+Au collisions. 


Radial dependence of the in-jet $p/\pi$ ratio is presented in Fig.~\ref{dR_fig} for charged-particle tracks with $2.0 < p_{\rm{T}} < 3.0$ GeV/$c$ in $R = 0.3$ jets with $p^{\rm{cons}}_{\rm{T}} > 2.0$ GeV/$c$, and $p^{\rm{jet}}_{\rm{T}} > 9.0$ GeV/$c$.
The systematic uncertainty on the $p/\pi$ ratio for this measurement has been separated into fully correlated and uncorrelated contributions to isolate the shape uncertainty in Au+Au from the total uncertainty, given that the predicted trend relies on shape, rather than overall magnitude. For the $\Delta r$ range studied, 
no statistically significant trend consistent with the predicted linear increase between Au+Au and $p$+$p$ collision results is observed within the current uncertainties.

\section{Summary and Conclusions}
\label{summ}
The first measurement of in-jet relative baryon-to-meson production from the STAR experiment is presented for $p$+$p$ and $0-10\%$ central Au+Au collisions at $\sqrt{s_{\mathrm{NN}}} = 200$ GeV. The study of the $p$+$p$ data sample has shown a strong preference for pion production relative to proton production for jets with $p^{\rm{jet}}_{\rm{T}} > 9.0$ GeV/$c$, $R = 0.2$--$0.4$, and both constituent $p^{\rm{cons}}_{\rm{T}}$ thresholds, $ 2.0$ and $3.0$ GeV/$c$. This ratio is lower than the inclusive ratio for $p$+$p$ events of the same energy. Measurements reported for $0  - 10 \%$ central Au+Au collisions with the same kinematic selection of jets manifest a similar preference for pion production relative to proton production among jet fragments compared to $p$+$p$ collisions. The measured in-jet $p/\pi$ ratios in $0  - 10 \%$ central Au+Au collisions are consistent with those in p+p collisions and remain significantly below the inclusive $p/\pi$ ratios from $0  - 20 \%$ Au+Au collisions. These observations suggest that, within the current kinematic selections, jet fragmentation retains a predominantly vacuum-like character. Further studies with extended kinematic reach and reduced selection biases will be essential to fully assess potential medium-induced modifications.
The in-jet $p/\pi$ ratio as a function of $\Delta r$ shows little deviation between the two systems, indicating no evidence of shower-thermal recombination for in-jet hadronization.

\section*{Acknowledgments}
We thank the RHIC Operations Group and SDCC at BNL, the NERSC Center at LBNL, and the Open Science Grid consortium for providing resources and support.  This work was supported in part by the Office of Nuclear Physics within the U.S. DOE Office of Science, the U.S. National Science Foundation, National Natural Science Foundation of China, Chinese Academy of Science, the Ministry of Science and Technology of China and the Chinese Ministry of Education, NSTC Taipei, the National Research Foundation of Korea, Czech Science Foundation and Ministry of Education, Youth and Sports of the Czech Republic, Hungarian National Research, Development and Innovation Office, New National Excellency Programme of the Hungarian Ministry of Human Capacities, Department of Atomic Energy and Department of Science and Technology of the Government of India, the National Science Centre and WUT ID-UB of Poland, German Bundesministerium f\"ur Bildung, Wissenschaft, Forschung and Technologie (BMBF), Helmholtz Association, Ministry of Education, Culture, Sports, Science, and Technology (MEXT), Japan Society for the Promotion of Science (JSPS), and Agencia Nacional de Investigacion y Desarrollo de Chile (ANID), Chile.




\bibliographystyle{unsrt}
\bibliography{refs}





\end{document}